\documentclass[aps,prb,preprint,superscriptaddress]{revtex4}

\usepackage{graphicx}
\usepackage{subfigure}
\usepackage{bm}    
\usepackage{amsmath}  
\usepackage{amssymb}
\usepackage{color}

\newcommand{\ket}[1]{|#1\rangle}

\begin{document}

\title{Subsystem constraints in variational second order density matrix optimization: curing the dissociative behavior}
\author{Brecht Verstichel}
\email{brecht.verstichel@ugent.be}
\affiliation{Ghent University, Center for Molecular Modeling, Proeftuinstraat 86, B-9000 Gent, Belgium}
\author{Helen van Aggelen}
\affiliation{Ghent University, Department of Inorganic and Physical Chemistry, Krijgslaan 281 (S3), B-9000 Gent, Belgium}
\author{Dimitri Van Neck}
\affiliation{Ghent University, Center for Molecular Modeling, Proeftuinstraat 86, B-9000 Gent, Belgium}
\author{Paul W. Ayers}
\affiliation{McMaster University, Department of Chemistry, Hamilton, Ontario, L8S 4M1, Canada}
\author{Patrick Bultinck}
\affiliation{Ghent University, Department of Inorganic and Physical Chemistry, Krijgslaan 281 (S3), B-9000 Gent, Belgium}

\date{\today}

\begin{abstract}
A previous study of diatomic molecules revealed that variational second-order density matrix theory has serious problems in the dissociation limit when the N-representability is imposed at the level of the usual two-index ($P$, $Q$, $G$) or even three-index ($T_1$, $T_2$) conditions [H. van Aggelen et al., Phys. Chem. Chem. Phys. 11, 5558 (2009)]. Heteronuclear molecules tend to dissociate into fractionally charged atoms. In this paper we introduce a general class of $N$-representability conditions, called subsystem constraints, and show that they cure the dissociation problem at little additional computational cost. As a numerical example the singlet potential energy surface of $\text{BeB}^+$ is studied. The extention to polyatomic molecules, where more subsystem choices can be identified, is also discussed.
\end{abstract}
\maketitle

\section{Introduction}
In recent years much attention has been devoted to the direct determination of the second-order density matrix (2DM) through variational optimization. As first shown by Husimi \cite{husimi}, the energy of a system interacting with at most two-particle interactions is fully determined by the 2DM. Some fifteen years later L\"owdin \cite{lowdin} independently derived similar results and suggested determining the 2DM directly in a variational approach. The first practical calculation was done by Mayer \cite{mayer} who tried to compute the energy of an electron gas by a variational optimization of the 2DM. The energies obtained however, were much too low, and inconsistent with existing results. Tredgold \cite{tredgold} realized that the problem arises because the set of density matrices over which the optimization is carried out is too large. Although in these first attempts some obvious constraints were included, better constraints are needed in order to make sure that the 2DM can be derived from a physical wavefunction. This problem was termed the $N$-representability problem by Coleman in his seminal review paper \cite{coleman}, in which he solved the ensemble $N$-representability problem for the first-order density matrix (1DM) and derived some bounds for the eigenvalues of the 2DM. Garrod and Percus \cite{garrod} subsequently derived the much stronger positivity conditions $Q$ and $G$. Because of the computational complexity and some dissapointing results on nuclei \cite{mihailovic}, not much progress was made the next twenty-five years. Interest renewed in the direct variational determination of the 2DM after Nakata et al. \cite{nakata_first} and then Mazziotti \cite{mazziotti} used a semidefinite program algorithm (SDP) to study a number of small atoms and molecules and got reasonably accurate results. These results sparked of a lot of developments. New $N$-representability conditions where introduced, e.g. the three-index $T$ conditions, as set forth by Zhao et al. \cite{zhao}, which led to mHartree accuracy \cite{hammond,nakata_last,mazz_T_con,Gido_T_con,mazz_book,braams_book} for molecules near equilibrium geometries, and generalizations thereof$~$\cite{maz_gen_T_1,maz_gen_T_2,maz_gen_T_3}. Algorithmic breakthroughs were realized with the implementation of a $r^6$ scaling SDP algorithm \cite{maz_prl,maz_alg_2,maz_alg_3} and the development of an active-space variational 2DM method \cite{maz_as_1,maz_as_2,greenman}. A drastic failure of the standard $N$-representability constraints ($PQGT$) was shown to occur in the dissociation limit by van Aggelen et al. \cite{helen_1} using a recently developed semidefinite programming code \cite{atomic}. In this article we propose new strict constraints, which we call subsystem constraints, that fix the inaccuracies in the dissociation limit. Sec. \ref{theory} contains the theoretical derivation of the subsystem constraints in a general framework. It is shown that identifying a subspace of the complete single-particle space leads to upper bounds for the energy of the total system, that must be obeyed by any $N$-representable 2DM. As a simple illustration, the technique is applied in Sec. \ref{application} to the dissociation of $\text{BeB}^+$ in a small (Dunning-Hay) basis set. Sec. \ref{conc} contains a summary and discussion. A systematic and thorough study of the diatomic potential energy surfaces for the 14-electron series is the subject of a separate publication \cite{helen_2}.

\section{\label{theory} Theory}
\subsection{Integer-$N$ ensemble representability}
The second-order density matrix (2DM) $\Gamma^N$ corresponding to an $N$-fermion wavefunction 
$|\Psi^N\rangle$ is defined as 
\begin{equation}
\Gamma^N_{\alpha\beta;\gamma\delta} =
\langle \Psi^N|a^\dagger_\alpha a^\dagger_\beta a_\delta a_\gamma |\Psi^N\rangle .  
\label{basic}
\end{equation}
Second-quantized notation is used where $a^\dagger_\alpha$ ($a_\alpha$) creates (annihilates) 
a fermion in the single-particle (sp) state $\alpha$. The sp basis is assumed to be orthonormal throughout the article. 
Eq.~(\ref{basic}) is easily generalized to the 2DM corresponding to an ensemble of $N$-fermion wavefunctions; 
conversely, a matrix $\Gamma^N$ is called 
integer-$N$ ensemble representable if 
\begin{equation}
\Gamma^N_{\alpha\beta;\gamma\delta} =\sum_i x_i 
\langle \Psi^N_i|a^\dagger_\alpha a^\dagger_\beta a_\delta a_\gamma |\Psi^N_i\rangle 
\end{equation}
for some set of $N$-fermion wave functions $|\Psi^N_i\rangle$ and positive weights $x_i$ obeying 
$\sum_i x_i =1$. 

We consider a system governed by a Hamiltonian containing a one-body part $t$ and a two-body interaction $V$, 
\begin{eqnarray}
\hat{H} &=& \sum_{\alpha\gamma}t_{\alpha\gamma}a^\dagger_\alpha a_\gamma + \frac{1}{4}\sum_{\alpha\beta\gamma\delta}V_{\alpha\beta;\gamma\delta}a^\dagger_\alpha 
a^\dagger_\beta a_\delta a_\gamma ,
\end{eqnarray}
where $V_{\alpha\beta ; \gamma\delta}$ represents the antisymmetrized matrix elements of the interaction. 
The exact ground-state energy $E^N_0$ (assumed to be nondegenerate) is determined by finding the 2DM $\Gamma^N$ that 
minimizes the energy functional 
\begin{eqnarray}
E^N_0 &=& \underset{\Gamma^N}{\min} \left[ \mbox{Tr}(t\rho^N) + \mbox{Tr}(V\Gamma^N)\right],\\
&&\left\{\begin{array}{l}
\rho^N = \frac{1}{N-1}\bar{\Gamma}^N\\
\mbox{$\Gamma^N$ is integer-$N$ ensemble representable}\end{array}\right. 
\nonumber
\end{eqnarray}
where $\Gamma^N$ is subject to the constraints listed after the $\{$-symbol. Note that 
the Tr(ace) operation in the 2DM space is restricted to antisymmetric two-index combinations and 
that the first-order density matrix $\rho^N$ is defined through the partial trace 
$\bar{\Gamma}_{\alpha\gamma} = \sum_\beta \Gamma_{\alpha\beta;\gamma\beta}$. 

A lower bound to the exact energy is obtained by the variational determination of the 2DM subject to a selected class of necessary $N$-representability conditions. This  
reduces to finding the 2DM $\Gamma^N$ that minimizes the energy functional  
\begin{eqnarray}
\label{int_N}
E^N_{SDP} &=& \underset{\Gamma^N}{\min} \left[ \mbox{Tr}(t\rho^N + \mbox{Tr}(V\Gamma^N)\right]\\ 
&&\left\{\begin{array}{l}
\rho^N = \frac{1}{N-1}\bar{\Gamma}^N\\
\mbox{Tr}\,\Gamma^N = \frac{1}{2}N(N-1)\\
\mathcal{L}_i (\Gamma^N, N) \geq 0 \end{array}\right. 
\nonumber
\end{eqnarray}
The  $\mathcal{L}_i (\Gamma, N)$ are matrix functionals of the 2DM which are required to be positive semidefinite; 
they reflect a choice of necessary conditions for integer-$N$ ensemble representability,   
e.g.\ the two-index $P$, $Q$ and $G$ condition or the three-index $T_1$ and $T_2$ conditions. As indicated in the notation, the matrix functionals also depend on the particle number.  
 
\subsection{\label{fracN}Fractional-$N$ ensemble representability}

The exact solution for a fractional electron number $\bar{N}$ has only one sensible 
definition, based on considering ensembles containing wave functions with various electron numbers, 
and the resulting energy then has the well-known piecewise linear behavior between integer values \cite{frac_N_1,frac_N_2,savin}.  

Reformulated in terms of density matrices, one defines a 2DM 
$\Gamma$ to be $\bar{N}$-representable if 
\begin{equation}
\Gamma = \sum_N x_N \Gamma^N
\end{equation}         
for a set of integer-$N$ ensemble representable $\Gamma^N$ and  
a set of positive weights $x_N$ obeying $\sum_N x_N =1$ and $\sum_N Nx_N = \bar{N}$. 

Obviously, the 1DM $\rho$ corresponding to the same ensemble cannot be obtained directly from 
$\Gamma$ by a partial-trace operation without knowledge of the ensemble weights. 
It is therefore more natural to consider the \textit{pair} $(\rho ,\Gamma)$ as being 
fractional-$\bar{N}$ ensemble representable if 
\begin{eqnarray}
\left\{\begin{array}{l}
\rho = \sum_N x_N \rho^N;\; \Gamma = \sum_N x_N \Gamma^N\\
x_N \geq 0 ;\; \sum_N x_N =1;\; \sum_N Nx_N =\bar{N}\\
\forall N: \rho^N = \frac{1}{N-1}\bar{\Gamma}^N\\
\forall N: \mbox{$\Gamma^N$ is integer-$N$ ensemble representable}
\end{array}\right. 
\end{eqnarray}
The exact solution is then simply generated by the minimization problem
\begin{eqnarray}
E^{\bar{N}}_0 &=& \underset{\rho ,\Gamma}{\min} \left[ \mbox{Tr}(t\rho) + \mbox{Tr}(V\Gamma)\right],\\
&&\left\{\begin{array}{l}
\mbox{$(\rho, \Gamma$) is fractional-$\bar{N}$ ensemble representable}\end{array}\right. 
\nonumber
\end{eqnarray}

It is now clear how the variational problem for a \textit{selected} choice of N-representability conditions 
[corresponding to Eq.~(\ref{int_N})] should be phrased, when it is generalized to a fractional electron number:    
one should minimize 
\begin{eqnarray}
E^{\bar{N}}_{SDP} &=& \underset{x_N ,\Gamma^N}{\min} \left[ \mbox{Tr}(t\rho) + \mbox{Tr}(V\Gamma)\right]
\label{eq10}\\ 
&&\left\{\begin{array}{l}
\rho = \sum_N x_N \rho^N; \Gamma = \sum_N x_N \Gamma^N\\
x_N \geq 0 ; \sum_N x_N =1; \sum_N Nx_N =\bar{N}\\
\forall N: \rho^N = \frac{1}{N-1}\bar{\Gamma}^N\\
\forall N: \mbox{Tr}\,\Gamma^N = \frac{1}{2}N(N-1)\\
\forall N: \mathcal{L}_i (\Gamma^N, N) \geq 0 \end{array}\right. 
\nonumber
\end{eqnarray}
where both the weights $x_N$ and the 2DM's $\Gamma^N$ can be varied. 
For any choice of weights $x_N$ the energy is minimal when 
$\Gamma^N$ corresponds to the SDP solution for integer $N$, so Eq.~(\ref{eq10}) can be reformulated as  
\begin{eqnarray}
E^{\bar{N}}_{SDP} &=& \underset{x_N}{\min} \sum_N x_N E^N_{SDP}\\
&&\left\{\begin{array}{l}
x_N \geq 0 ; \sum_N x_N =1; \sum_N Nx_N =\bar{N}\end{array}\right.
\nonumber
\end{eqnarray}
This leads naturally to a piecewise linear solution \cite{frac_N_1} which, for a convex set $E^N_{SDP}$, is given by 
\begin{equation}
E^{\bar{N}}_{SDP}=
(\bar{N}-\mathrm{Int}(\bar{N}))E^{\mathrm{Int}(\bar{N}) +1}_{SDP}+(\mathrm{Int}(\bar{N}) +1 -\bar{N})E^{\mathrm{Int}(\bar{N})}_{SDP}
\end{equation} 
where the function $\mathrm{Int}(x)$ returns the nearest integer number smaller than $x$. So in exactly the same way as for the exact solution, 
nothing new emerges from introducing fractional electron numbers. 
There is, however,  one interesting observation, that is crucially important for 
the subsequent derivation of subsystem constraints: 
for any Hamiltonian $(t,V)$ the energy evaluated with a fractional-$\bar{N}$ ensemble representable density 
$(\rho ,\Gamma)$ obeys the inequalities
\begin{equation}
E(\rho ,\Gamma)\geq E^{\bar{N}}_0\geq E^{\bar{N}}_{SDP}. 
\label{equ12}
\end{equation}
The first inequality follows from the definition~(\ref{eq10}) of $E^{\bar{N}}_0$ as the minimum 
over ensembles, the second one from the fact that for all integer $N$, $E^N_{SDP}$ is a lower bound 
for $E_0^N$. 

Some electronic structure methods (like Hartree-Fock or density functional theory) can be easily generalized to accomodate a fractional number of electrons.  In that case the behavior of the energy in between integer values usually disagrees with the piecewise linear result from an exact calculation \cite{cohen,mori,ruzs,perdew,morisa,ajcoh,arxiv}.

Something similar is also present in the SDP technique.  A naive extension of this framework to fractional electron number would be to treat the parametric dependence on $N$ in Eq.~(\ref{int_N}) as a continuous variable: this is immediately understood to be unphysical, as no reference is made to the ensemble interpretation. However, as first noted by van Aggelen et. al. \cite{helen_1}, SDP computations on the union of isolated subsystems become equivalent to allowing the number of electrons on each subsystem to be a continuous variable, again without reference to the ensemble. Application of the SDP to the dissociation limit of diatomic molecules and ions results in dissociated atoms with fractional occupancies in the case of a heteronuclear diatomic, which is chemically clearly unacceptable. In order to prevent this, new $N$-representability conditions are needed.

\subsection{\label{subcon} Subsystem constraints}

As before, Greek indices $\alpha,\beta,..$ are used for the full set of single-particle states. We now introduce an (arbitrary) subset;  Roman indices $a,b,..$ are used if we want to restrict the orbitals to members of this subset. In a polyatomic molecule, e.g., we may consider an orthonormal basis of the subspace generated by the basisfunctions centered on a particular atom. This selection of a subset is in fact equivalent to choosing a subspace of the complete single-particle space, since the SDP setup with the standard two-index or three-index conditions (of the $P$, $Q$, $G$, $T_1$, $T_2$ type) is invariant under a unitary transformation in \textit{sp} space. Note that this invariance is broken when in a particular \textit{sp} basis only diagonal constraints are kept \cite{ww,aydav,davidson_2,mcrea,davidson_1,nak_diag,hammond,kamarchik}. 

The derivation of the subsystem constraints starts by noting that, if $\Gamma^N$ is integer-$N$ ensemble representable, then the pair $(\rho^\text{sub},\Gamma^\text{sub})$,
\begin{eqnarray}
\Gamma^{\text{sub}}_{ab;cd} &=& \Gamma^N_{ab;cd}\\
\rho^{\text{sub}}_{ac}&=&\frac{1}{N-1}\sum_\beta \Gamma^N_{a\beta;c\beta}=\rho^N_{ac} 
\end{eqnarray}
is fractional-$\bar{N}$ ensemble representable in the Fock space generated 
by the subset, with $\bar{N} = \sum_a \rho^{\text{sub}}_{aa}$. To prove this we start with the integer-$N$ ensemble representability of $\Gamma^N$:
\begin{equation}
\label{gamma_sub}
\Gamma^\text{sub}_{ab;cd} = \sum_i x_i \langle\Psi^N_i|a^\dagger_a a^\dagger_b a_d a_c  |\Psi^N_i\rangle. 
\end{equation}  
We can expand each $|\Psi^N_i\rangle$ in Slater determinants, classified according to the number of subsystem orbitals they contain:
\begin{equation}
|\Psi^N_i\rangle = \sum_{j s_j \bar{s}_{N-j}} C_{ijs_j\bar{s}_{N-j}} |s_j \bar{s}_{N-j}\rangle 
\end{equation}
in which $0\leq j\leq N$, $s_j$ represents a set of $j$ subsystem orbitals, and $\bar{s}_{N-j}$ a set of $N-j$ orbitals not in the subsystem. 
Using the fact that the string of subsystem-type creation/annihilation operators in Eq. (\ref{gamma_sub}) does not change the number of subsystem orbitals, that it leaves the non-subsystem part of the Slater determinant unchanged, and using orthonormality of the $\bar{s}_{N-j}$ states, we see that 
\begin{equation}
\sum_i x_i \langle\Psi^N_i|a^\dagger_a a^\dagger_b a_d a_c  |\Psi^N_i\rangle = \sum_i x_i \sum_{j \bar{s}_{N-j}} \langle\Psi^{j}_{i\bar{s}_{N-j}}|a^\dagger_a a^\dagger_b a_d a_c  |\Psi^{j}_{i\bar{s}_{N-j}}\rangle~,
\end{equation}  
where 
\begin{equation}
|\Psi^{j}_{i\bar{s}_{N-j}}\rangle  = \sum_{s_j} C_{ijs_j \bar{s}_{N-j}}
|s_j\rangle
\end{equation}
is a state with $j$ particles in the Fock space generated by the subsystem orbitals. These states are obviously not normalized, their norm is given by 
\begin{equation}
\langle\Psi^{j}_{i\bar{s}_{N-j}}|\Psi^{j}_{i\bar{s}_{N-j}}\rangle = \sum_{s_j} |C_{ijs_j \bar{s}_{N-j}}|^2 = w^j_{i\bar{s}_{N-j}} ~.
\end{equation}
If we replace them by normalized states
\begin{equation}
|\tilde{\Psi}^{j}_{i\bar{s}_{N-j}}\rangle  = [w^j_{i\bar{s}_{N-j}}]^{-1/2} |\Psi^{j}_{i\bar{s}_{N-j}}\rangle ~,
\end{equation}
it follows that 
\begin{equation}
\Gamma^\text{sub}_{ab;cd} = \sum_{j;i\bar{s}_{N-j}} x_i w^j_{i\bar{s}_{N-j}} \langle\tilde{\Psi}^{j}_{i\bar{s}_{N-j}}|a^\dagger_a a^\dagger_b a_d a_c  |\tilde{\Psi}^{j}_{i\bar{s}_{N-j}}\rangle ~,
\end{equation}
where 
\begin{equation}
\sum_{j;i\bar{s}_{N-j}} x_i w^j_{i\bar{s}_{N-j}}=1~,
\end{equation}
because of the normalization of the original $N$-particle states. 
In an analogous way one shows that the first-order density matrix $\rho^\text{sub}$ can be written as
\begin{equation}
\rho^\text{sub}_{ac} = \sum_{j;i\bar{s}_{N-j}} x_i w^j_{i\bar{s}_{N-j}} \langle\tilde{\Psi}^{j}_{i\bar{s}_{N-j}}|a^\dagger_a a_c  |\tilde{\Psi}^{j}_{i\bar{s}_{N-j}}\rangle  ~.
\end{equation}
This proves that $\Gamma^\text{sub}$ and $\rho^\text{sub}$ can be derived from the same ensemble of wave functions containing only orbitals in the subsystem. 
This ensemble has a fractional number of particles (in the subsystem space) given by 
$\bar{N}= \sum_{j;i\bar{s}_{N-j}}j x_i w^j_{i\bar{s}_{N-j}}= \sum_a \rho^\text{sub}_{aa}$.
 
Based on Eq.~(\ref{equ12}), the following necessary condition then holds: 
If $\Gamma^N$ is integer-$N$ ensemble representable, then $(\rho^\text{sub} ,\Gamma^\text{sub})$ should obey 
\begin{equation}
\label{24}
\mbox{Tr}\left(t^\text{sub}\rho^\text{sub}\right) +\mbox{Tr}\left(V^{\text{sub}}\Gamma^{\text{sub}}\right) \geq E^{\bar{N}}_{SDP}
\end{equation}
with $\bar{N}=\sum_a \rho^\text{sub}_{aa}$, for any Hamiltonian $(t^\text{sub} ,V^\text{sub})$ defined 
in the subspace. These subsystem constraints can be quite powerful; they are grossly violated in Coulombic dissociation problems as documented in \cite{helen_1}. Note that for consistency it is preferable to use the SDP lower bound for the subsystem energy rather than the exact one, even if this should be available.

The subsystem constraints in Eq.~(\ref{24}) were derived for a particular basis choice of the subspace, \textit{i.e.} a subset of the underlying 
orthonormal basis $|\alpha\rangle$ of the total single-particle space. However, the subsystem constraints only depend on the subspace itself, which is most easily seen 
by 
extending the $t^\text{sub}$ and $V^\text{sub}$ operators (defined on the subspace) to the total single-particle space using projection operators.  
One can rewrite, e.g., 
\begin{equation}
\mbox{Tr}\left(t^\text{sub}\rho^\text{sub}\right) = \sum_{ac}t^\text{sub}_{ac}\rho^\text{sub}_{ac}= 
\sum_{\alpha\gamma}\rho^N_{\alpha \gamma} t^\text{sub}_{\alpha \gamma},
\end{equation}
where 
\begin{equation}
t^\text{sub}_{\alpha\gamma} = \langle \alpha | \tilde{P}\tilde{t}^\text{sub}\tilde{P}|\gamma\rangle . 
\end{equation} 
Here $\tilde{t}^\text{sub}$ and $\tilde{P}$ are first-quantized operators with 
\begin{equation}
\tilde{P} = \sum_a |a\rangle \langle a|
\end{equation}
the orthogonal projector onto the subspace. 
In the same way  one can rewrite the two-body operator defined on the subspace as 
\begin{equation}
\mbox{Tr}\left(V^\text{sub}\Gamma^\text{sub}\right) = \frac{1}{4}\sum_{abcd}\Gamma^\text{sub}_{ab;cd}V^\text{sub}_{ab;cd}=
\frac{1}{4}\sum_{\alpha\beta\gamma\delta}\Gamma^N_{\alpha\beta ;\gamma\delta} V^\text{sub}_{\alpha\beta ;\gamma\delta} 
\end{equation}
by defining 
\begin{equation}
V^\text{sub}_{\alpha\beta ;\gamma\delta} = \langle \alpha \beta | \tilde{P}_1\tilde{P}_2 \tilde{V}^\text{sub}\tilde{P}_2\tilde{P}_1 
|\gamma \delta \rangle .
\end{equation}

\subsection{Implementation for diatomics}

For a diatomic molecule the molecular $SDP$ solution tends to localize a fractional number of electrons on a well-separated atom, 
whenever this situation is energetically favorable in the continuous-$N$ sense mentioned at the end of Sec. \ref{fracN}. As a result, 
the energy of the atomic subsytem drops below the true (ensemble-based) energy and a violation of the inequality (\ref{24}) occurs.
In this case the choice of the subsystems is therefore obvious: in the dissociation limit they should coincide with the individual atoms. 
While a general basis-set independent formulation is possible, in practice the calculations are performed using atom-centered basis functions 
for which this requirement is automatically satisfied. 

The procedure for applying the atom-$A$ subsystem constraint in a diatomic $AB$ can then be summarized as follows: 
\begin{enumerate}
\item Solve the atomic $SDP$ problem for a central charge $Z_A$ at various electron numbers, using the \textit{sp} orbitals centered on $A$.
In practice, only electron numbers near atomic neutrality are important. This generates the atomic energy $E^{\bar{N}}_A$ as a function of fractional 
electron number $\bar{N}$ (i.e. a piecewise linear curve). 
\item For each internuclear distance $R_{AB}$, calculate the transformation matrix between the (nonorthogonal) atomic basis of the $A$-centered orbitals $\ket{i_A}$, and the orthonormal molecular basis $\ket{\alpha}$ that is used in the SDP program,
\begin{equation}
U^A_{\alpha ,i} = \langle i_A|\alpha\rangle.
\end{equation}  
The $U^A$ matrix is easily constructed with standard quantities in molecular modelling packages, 
\begin{equation}
U^A_{i,\alpha} = \sum_{j_D} C_{\alpha ,j_D} S_{i_A;j_D}
\end{equation}
with $C$ the expansion coefficients of the $|\alpha\rangle$ molecular basis in terms of all the nonorthogonal orbitals centered on the various atoms,
\begin{equation}
|\alpha\rangle = \sum_{j_D} C_{\alpha ; j_D}|j_D\rangle ,
\end{equation}
and with $S_{i_A ;j_D} = \langle i_A|j_D \rangle$ the overlap matrix for the atom-centered basis functions. 
The orthogonal projector onto the subspace spanned by the $A$-centered orbitals, when expressed in terms of the nonorthogonal basis set, reads \cite{bultinck}
\begin{equation}
\tilde{P}^A = \sum_{ij}(S^{-1}_A)_{ij}|i_A\rangle\langle j_A|  
\end{equation}
where $S_A$ is the block of the overlap matrix corresponding to the $A$-centered orbitals, and $S_A^{-1}$ is the inverse of this block.
\item Perform the SDP program with the extra linear inequality: 
\begin{equation}
\label{26}
\mbox{Tr}~\left(t^A\rho^N\right)+\mbox{Tr}~\left(V^A\Gamma^N\right) \geq E_A^{\bar{N}=\mbox{Tr}\left(1^A\rho^N\right)}.
\end{equation}   
Here: 
\begin{eqnarray}
(t^A)_{\alpha\gamma} &=& \sum_{ij}\langle i_A |\tilde{t}^A| j_A\rangle W^A_{i\alpha}W^A_{j\gamma}\\
(1^A)_{\alpha\gamma} &=& \sum_{ij} (S_A)_{ij}  W^A_{i\alpha}W^A_{j\gamma}\\
(V^A)_{\alpha\beta;\gamma\delta} &=& \sum_{ijkl} \langle i_A j_A|\tilde{V}^A|k_Al_A\rangle 
W^A_{i\alpha}W^A_{j\beta}W^A_{k\gamma}W^A_{l\delta}
\end{eqnarray}
and $t_A$ is the kinetic energy plus attraction to nucleus $A$. The coefficients $W$ are given by:
\begin{equation}
\label{W}
W^A_{i\alpha} = \sum_j U^A_{\alpha j}\left(S_A^{-1}\right)_{ji}
\end{equation}
The inequality (\ref{26}) is nothing but the application of Eq. (\ref{24}) in the subspace 
defined by the \textit{sp} orbitals centered on $A$ and using the Hamiltonian of atom $A$. 
Obviously, atom $B$ generates a similar inequality.
\end{enumerate}
The inequalities for $A$ and $B$ are expected to become important in the dissociation limit, as they prevent an artificial lowering of the energy due to fractional 
electron numbers on atom $A$ and $B$.

\section{\label{application}Numerical verification}

It has been found previously that the 2DM variational optimization under $P$, $Q$ and $G$ constraints leads to chemically flawed dissociation curves where the atoms carry noninteger numbers of electrons even when separated by a large distance. This is especially true for molecular ions and persists even when including $T$ constraints \cite{helen_1}. In order to show the value of the subsystem constraints, we present the potential energy surface of $\text{BeB}^+$, computed for a separation ranging from 1 to 9 \AA. The main interest here is a proof-of-principle of the fact that the new constraints indeed severely restrict the variational freedom in the SDP. We therefore opted for the fairly small Dunning-Hay basis \cite{dh}, making full-CI calculations still feasible. $\text{BeB}^+$ is a good example since this 8 electron system dissociates into $\text{Be}$ and $\text{B}^+$. Application of $P$, $Q$ and $G$ for an 8 electron system is expected to yield energies that are significantly too low compared to full-CI. However, at the dissociation limit, the energy of the molecule should be equivalent to that of isolated $\text{Be}$ and $\text{B}^+$. As shown in \cite{atomic}, for both $\text{Be}$ and $\text{B}^+$ the $P$, $Q$ and $G$ energy is very nearly equal to the full-CI energy, so application of the subspace constraints should result in much higher $PQG$ energies than when not using the subspace constraints.

\begin{center}
\begin{figure}
\includegraphics[scale=0.6]{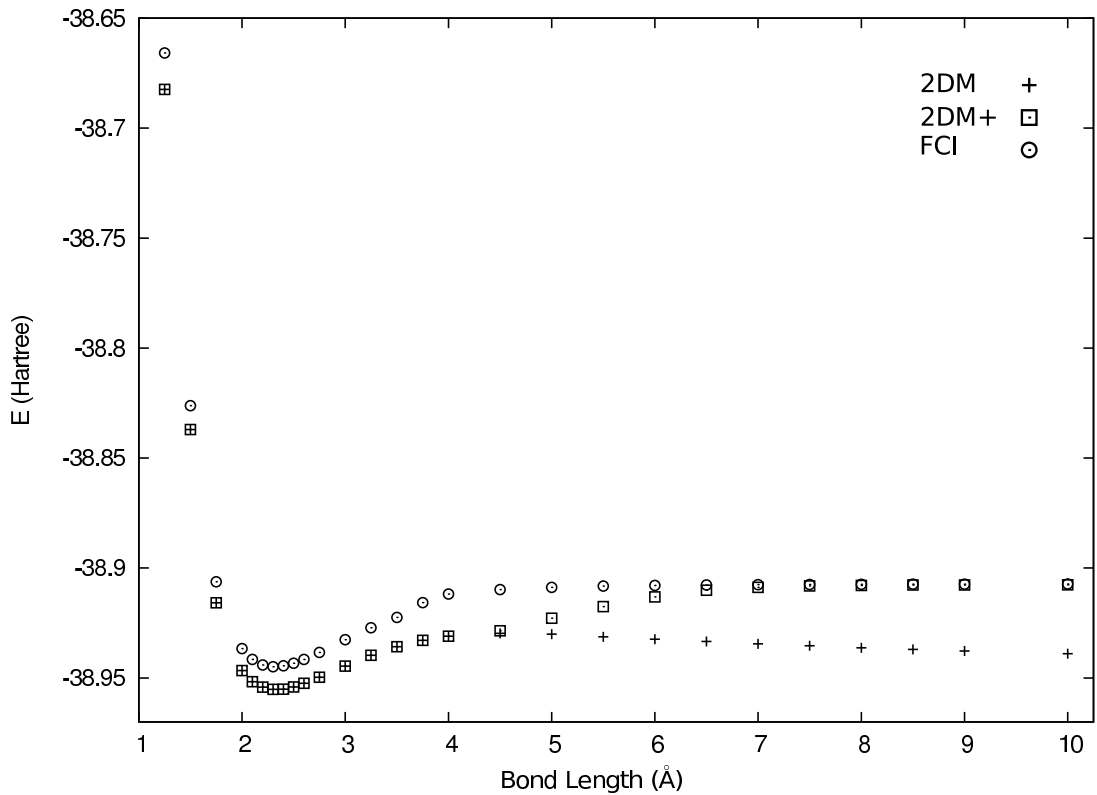}
\caption{\label{BeB_p} The singlet dissociation energy curve of $\text{BeB}^+$ calculated in full-CI, and determined variationally without (2DM) and with (2DM+) subspace constraints.}
\end{figure}
\end{center}

Table \ref{BeB_t} shows the energy of the molecule at different internuclear distances, computed at the full-CI level of theory as well as the variationally optimized 2DM energy with and without subspace constraints (see also Fig. \ref{BeB_p}).
\begin{table}
\caption{\label{BeB_t}Difference (in mHartree) between full-CI energy (FCI) and variationally optimized 2DM energy without (2DM) and with (2DM+) subspace constraints.}
\begin{ruledtabular}
\begin{tabular}{ccccccc}
$R$&2DM&2DM+&\vline&R&2DM&2DM+\\
\hline
1.25  &  16.55 &  16.55 &\vline&3.50  &  13.30 &  13.30 \\
1.50  &  10.86 &  10.86 &\vline&3.75  &  17.17 &  17.17 \\
1.75  &   9.56 &   9.56 &\vline&4.00  &  19.12 &  19.12 \\
2.00  &   9.94 &   9.94 &\vline&4.50  &  19.90 &  18.73 \\
2.10  &  10.07 &  10.07 &\vline&5.00  &  21.28 &  14.07 \\
2.20  &  10.07 &  10.07 &\vline&5.50  &  23.07 &   9.35 \\
2.30  &  10.30 &  10.30 &\vline&6.00  &  24.48 &   5.27 \\
2.40  &  10.59 &  10.59 &\vline&6.50  &  25.67 &   2.44 \\
2.50  &  10.77 &  10.77 &\vline&7.00  &  26.95 &   1.26 \\
2.60  &  10.91 &  10.91 &\vline&7.50  &  27.90 &   0.66 \\
2.75  &  11.30 &  11.30 &\vline&8.00  &  28.88 &   0.53 \\
3.00  &  12.00 &  12.00 &\vline&8.50  &  29.63 &   0.38 \\
3.25  &  12.56 &  12.56 &\vline&9.00  &  30.39 &   0.37
\end{tabular}
\end{ruledtabular}
\end{table}

Table \ref{BeB_t} very clearly shows that the difference between the FCI and 2DM energies is substantial when using only the $P$, $Q$ and $G$ constraints. Especially at longer separations the difference between full-CI and 2DM energies can amount to roughly 0.03 Hartree. The subspace constraints succeed in reducing this error by approximately two orders of magnitude. As expected, the remaining error is very small because $P$, $Q$ and $G$ yield energies for the atomic 4-electron isoelectronic series that are very near to full-CI energies. The present new constraints are clearly very succesful. As Figure \ref{BeB_p} shows, the constraints are active most for separations above 4.5~\AA. The nearer to complete dissociation, the more of the error is recovered by the subspace constraints. As shown by Van Aggelen et al. \cite{helen_2}, not only are the energies improved, also chemical observables and chemical concepts are substantially better for the 2DM obtained when including the subspace constraints. As an example, the Mulliken population \cite{mulliken} on the Be atom at $9~\text{\AA}$ is $+0.38$ when not using the subspace constraints, whereas inclusion of the subspace constraints yields a charge of $0.00$, consistent with what it should be according to the full-CI data.
The added constraints result in a much better description of molecular dissociation. Neither atom still suffers from fractional occupancy at the dissociation limit. Addition of each subsystem constraint does not slow down the SDP, as it adds a fairly simple linear inequality constraint. 
\section{\label{conc}Summary and conclusions}      
In this paper we derived necessary conditions that substantially reduce the dissociation problem found in diatomic molecules at large separation \cite{helen_1}. These conditions are based on fractional-$N$ ensemble representability. For any subset of single-particle space one can associate a "subsystem" 2DM and 1DM, which must be fractional-$N$ ensemble representable. Any Hamiltonian defined on the subspace leads to linear inequalities for the 2DM of the full system. In the case of diatomic molecules we can associate the subsystems with the single atoms and consider the separate atomic Hamiltonians. Application to $\text{BeB}^+$ shows that these constraints are strongly violated in the dissociation limit. When they are imposed during the optimization, they lead to a significant improvement in the energy, and cure the pathological behavior reported in \cite{helen_1}. In the dissociation limit of heteronuclear diatomic molecules the occupation numbers of the single atoms are now integer, as they should be. 

For polyatomic molecules, the most relevant choice of the subsystems is not so clear cut. If all possible combinations (mono-, di-, tri-, \ldots atomic subsystem) are included the number of constraints can grow quite big, and for each subsystem one needs to perform separate 2DM optimizations. For mono-atomic subsystems this is a one-time task for a given basis set as all the required Hamiltonians, energies etc. can be kept stored. However, for larger (di-,tri-,\ldots) subsystems the problem is that the required data are geometry dependent. As a consequence, for every molecular calculation one will need to also obtain the variationally optimized 2DM energies for all non mono-atomic subspaces. As 2DM optimizations are quite time consuming, this may add a lot of overhead time to a molecular 2DM optimization. Furthermore, the number of constraints to be included, and as such the number of extra 2DM energy calculations for the subspaces, grows rapidly. 
For a diatomic, 2 subspace constraints are needed, as illustrated above for $\text{BeB}^+$. For a molecule with 5 atoms, the number of constraints equals already 30 of which 25 need to be computed specifically for the molecular geometry considered. A further drawback is that it is sometimes hard to predict whether the constraint needed will involve a cationic or anionic subsystem and as a consequence, both are preferably included. It is very unlikely that all of these constraints are needed and only few will be violated when not included. Unfortunately, at this moment it is not yet clear how to decide a priori which are the most important constraints to be included.  Despite these drawbacks, these constraints are very strong and their inclusion highly desirable in the SDP. This is shown in Ref. \cite{helen_2} where the subsystem constraints have been used in a study of the 14-electron isoelectronic series.

We would like to stress that subsystem constraints are much more general then the atom-based constraints discussed in this paper. They hold for any subspace, e.g. also for an arbitrary selection of molecular orbitals (which may correspond to an active space around the Fermi-energy or a restricted set of orbitals of a certain symmetry). Of course, in general many constraints of this type will not be active. It will be interesting to see whether other systems can be found for which the subsystem constraints lead to improvements in the quality of the variationally obtained 2DM.
\section{Acknowledgements}
We gratefully acknowledge financial support from FWO-Flanders and the research council of Ghent University. P.B. acknowledges Andreas Savin and Paola Gori-Giorgi for fruitful discussions. P.W.A. acknowledges support from NSERC and Sharcnet. B.V., H.V.A., P.B. and D.V.N. are Members of the QCMM alliance Ghent-Brussels.
\bibliography{qsep.bib}
\bibliographystyle{unsrt}

\end{document}